\documentclass[aps,pre,preprint,groupedaddress,nofootinbib,showpacs,onecolumn]{revtex4-1}
\usepackage[english]{babel}
\usepackage{bbm}
\usepackage{amssymb}
\usepackage{amsmath,amsfonts,amsthm,bm} 
\usepackage{mathtools}
\usepackage{times}
\usepackage{bm}
\usepackage{comment}
\usepackage{hyperref}
\usepackage{natbib}
\usepackage{url}
\usepackage{xcolor}
\usepackage{multirow}
\usepackage{tabularx, ragged2e, booktabs}
\newcolumntype{Y}{>{\centering\arraybackslash}X}
\usepackage{adjustbox}
\usepackage{cprotect}

\begin{document}

\title{Beyond Forcing Scenarios: Predicting Climate Change through Response Operators in a Coupled General Circulation Model}
\author{Valerio Lembo$^{1,}$, Valerio Lucarini$^{1,2,3,}\footnote{\texttt{v.lucarini@reading.ac.uk}}$, Francesco Ragone$^{4,}$}
\affiliation{\small [1] Meteorologisches Institut, Universit\"at Hamburg, Hamburg, Germany}
\affiliation{\small [2] Department of Mathematics and Statistics, University of Reading, Reading, UK}
\affiliation{\small [3] Centre for the Mathematics of Planet Earth, University of Reading, Reading, UK}
\affiliation{\small [4] Laboratoire de Physique, ENS de Lyon, Lyon, France}
\date{\today}
\begin{abstract}
Global Climate Models are key tools for predicting the future response of the climate system to a variety of natural and anthropogenic forcings. Here we show how to use statistical mechanics to construct operators able to flexibly predict climate change for a variety of climatic variables of interest. We perform our study on a fully coupled model - MPI-ESM v.1.2 - and for the first time we prove the effectiveness of response theory in predicting future climate response to CO$_2$ increase on a vast range of temporal scales, from inter-annual to centennial, and for very diverse climatic quantities. We investigate within a unified perspective the transient climate response and the equilibrium climate sensitivity and assess the role of fast and slow processes. The prediction of the ocean heat uptake highlights the very slow relaxation to a newly established steady state. The change in the Atlantic Meridional Overturning Circulation (AMOC) and of the Antarctic Circumpolar Current (ACC) is accurately predicted. The AMOC strength is initially reduced and then undergoes a slow and  partial recovery. The ACC strength initially increases due to changes in the wind stress, then undergoes a slowdown, followed by a recovery leading to a overshoot with respect to the initial value. Finally, we are able to predict accurately the temperature change in the {\color{black}{Northern Atlantic.}}
\end{abstract}

\maketitle

\section{Introduction}
Climate change is arguably one of the greatest contemporary societal challenges \cite{Palmer24390} {\color{black}and one of the grand contemporary scientific endeavours \cite{Ghil2019}.} The provision of new and efficient ways to understand its mechanisms and predict its future development is one of the key goals of climate science. Global climate models (GCMs) are currently the most advanced tools for studying future climate change; their future projections are key ingredients of the reports of the Intergovernmental Panel on Climate Change (IPCC) and are key for climate negotiations \cite{IPCC2013}. 
For IPCC-class GCMs, future climate projections are usually constructed by defining a few climate forcing scenarios, given by changes in the composition of the atmosphere and in the land use, each corresponding to a different intensity and time modulation of the equivalent anthropogenic forcing. Typically, for each scenario an ensemble of simulations is performed, with each member differing in terms of initial conditions, applied forcing or chosen physical parametrizations. Subsequent phases of the Coupled Model Intercomparison Project (CMIP, {\color{black}{currently}} the sixth phase CMIP6 is active \cite{Eyring2016}), which is part of the Program for Climate Model Diagnosis and Intercomparison (PCMDI), allowed to define standardized experimental protocols for numerical simulations with several GCMs and for the evaluation of the model runs \cite{Eyring2016b,Lembo2019}. A bottleneck of this approach is that the consideration of an additional forcing scenario requires running a new ensemble of simulations. Additionally, for each forcing scenario, it is hard to disentangle the impact of each component of the forcing, e.g. different greenhouse gases with their concentration pathways and land surface alterations in geographically distinct regions. Finally, no rigorous prescription exists for translating the climate change projections if one wants to consider different time modulations of a given forcing pathway, e.g. a faster or slower CO$_2$ increase. 

\subsection{Response theory and climate change}\label{responseclimatechange}

A possible strategy to deliver flexible and accurate climate change projections is the construction of response operators able to transform inputs given by forcing scenarios into outputs in the form of climate change signal. In this regard, it is tempting to use the fluctuation-dissipation theorem (FDT) \cite{Kubo1957,Bettolo2008}, which provides a dictionary for translating the statistics of free fluctuations of a system into its response to external forcings. The idea of using the FDT to predict climate change from climate variability has been first proposed by {\color{black}{Leith}} \cite{Leith1975} and used by several authors thereafter \cite{Alexeev2003,Cionni2004,Gritsun2007}.  The FDT has recently been key to inspiring the theory of emergent constraints, which are tools for reducing the uncertainties on climate change by looking at empirical relations between climate response and variability of some given observables \cite{Cox2018,Cox2019}.

In the case of nonequilibrium systems, forced fluctuations contain features that are absent from the free fluctuations of the unperturbed system \cite{Ruelle2009}. Therefore, the applicability of the FDT for such systems  faces some serious theoretical and practical challenges \cite{Lucarini2009,Gritsun2017,Lucarini2016,Ghil2019}. The climate is a nonequilibrium system whose dynamics is primarily driven by the heterogeneous absorption of solar radiation. The motions of the geophysical fluids with the associated transports of mass and energy, as well as the exchanges of infrared radiation, tend to re-equilibrate the system and allow it to reach a steady state \cite{Peixoto1992,Lucarini2014,Ghil2019}. Since the climate is not {\color{black}{in}} equilibrium, climate change projects only partially on the modes of climate variability, whilst climate surprises - unprecedented events - are indeed possible when forcings are applied \cite{Gritsun2017}. {\color{black}See Ghil and Lucarini \cite{Ghil2019} for a comprehensive mathematical and physical discussion of the relationship between climate variability and climate response to forcings.}

Response theory is a generalisation of the FDT that allows one to   to predict how the statistical properties of general systems - near or far from equilibrium, deterministic or stochastic - change as a result of forcings. After the pioneering work by {\color{black}{Kubo}} \cite{Kubo1957}, response theory has been firmly grounded in mathematical terms for stochastic \cite{Hairer2010} and deterministic \cite{Ruelle1998a,Ruelle1998b,Ruelle2009} systems. 
The use of response theory for predicting climate change has been successful in various numerical investigations performed on models of various degrees of complexity, ranging from rather simple ones \cite{Lucarini2009}, to intermediate complexity ones \cite{Abramov2008,Lucarini2011,Gritsun2007}, up to simple yet Earth-like climate models \cite{Ragone2016,Lucarini2016}. 
The key step is the computation of the Green function for each observable of interest. Then, the corresponding climate change signal is predicted by convolving  the Green function with the temporal pattern of forcing. Once the Green function is known, response theory allows one to treat in a unified and comprehensive way forcings with any temporal modulation, ranging from instantaneous to adiabatic changes. 

{\color{black}Note that, despite non-equilibrium conditions, a subtle relation exists between climate response and climate variability. Indeed, natural modes of variability of the unperturbed system that can be identified as prominent features in the autocorrelation or power spectra of climatic fields are described by the Ruelle-Pollicott poles \cite{Chekroun2014,Lucarini2018JSP,Ghil2019}. Such poles are responsible for extremely amplified response of the system to a resonating forcing with suitably defined spatial and temporal pattern.}  


%

{\color{black} A similar {\color{black} heuristic} approach, although not rooted in formal Ruelle's \cite{Ruelle1998a,Ruelle1998b,Ruelle2009} response theory, {\color{black} was also} proposed in the seminal work of Hasselmann \cite{Hasselmann1993}, which raised some concerns regarding its general applicability \cite{Good2011}. Despite this, recent attempts showed excellent skills for a wide range of observables in GCM experiments \cite{Good2011}. The crucial point is that in these works linear response formulas were applied to the output of individual experiments with GCMs. Ruelle's response theory clarifies that the heuristic idea of Hasselmann \cite{Hasselmann1993} is instead mathematically grounded if one considers ensemble averages rather than individual experiments.}

The climate models used so far to test response theory {\color{black}as formulated above} \cite{Ragone2016,Lucarini2016} lacked an active and dynamic ocean, so that the multiscale nature of climate processes was only partially represented. Capturing the slow oceanic processes is essential for a correct representation of the multidecadal and long-time climatic response. Encouragingly, response theory has recently been shown to have a great potential for predicting climate change in multi-model ensembles of CMIP5 atmosphere-ocean coupled GCMs outputs \cite{Aengenheyster2018}. Blending together data coming from different models is outside response theory theoretical framework, yet heuristically justifiable. However, a proper treatment within the boundaries of the theory, based on ensemble of simulations with the same model featuring an active ocean component, is still lacking.

{\color{black}{The response of a a slow (oceanic) climatic observable of interest {\color{black}has been investigated so far in relation to the} change in the dynamical properties of some other climatic observable {\color{black}, applying a}  convolution product \cite{Pillar2016,Kostov2017,Johnson2018,Cornish2020}. The strategy relies on constructing a linear regression between the predictand and predictor using the properties of the natural variability of the system. While indeed attractive and promising (and showing a good degree of success), this point of view cannot be rigorously ported to climate change studies. Using the resulting transfer function to predict the change of the observable of interest in a forced experiment would implicitly assume the validity of the FDT; see discussion above.}} 

{\color{black} We remark that, in some cases (but not always), it is theoretically possible to use  a climatic observable for predicting the response of another climatic observable of interest in the presence of an external forcing acting on the system. The possibility of using an observable as a surrogate forcing able to retain predictive power depends of non-trivial properties of its frequency-dependent response (its susceptibility, see below for a definition) and relies, in more practical terms, on the possibility of separating two or more dynamically relevant time scales in the system \cite{Lucarini2016b}. See recent applications of this idea in {\color{black}{Zappa et al.}} \cite{Zappa2020} and, using a methodology based on  adjoint modelling, in {\color{black}{Smith et al.}} \cite{Smith2019}.} 

\subsection{{\color{black}Predicting Climate Change using the Ruelle Response Theory}}
In this paper we show for the first time how {\color{black}{the Ruelle's response theory}} can be used to perform {\color{black}{successful centennial}} climate predictions in the fully coupled climate model MPI-ESM v.1.2 \cite{Giorgetta2013}.  

We consider two ensemble experiments. One experiment features an instantaneous CO$_2$ doubling  ($2xCO2$), and the outputs of its ensemble members are used to compute the Green functions of several observables of interest, as described in the \hyperref[met]{Methods} section (Appendix \ref{met}). We then perform an ensemble of runs where the CO$_2$ concentration is increased at the rate of 1\% per year until doubling ($1pctCO2$). We use the Green function computed with $2xCO2$ to reconstruct the response of $1pctCO2$ 
and compare the prediction with direct numerical simulations.
As discussed in the \hyperref[met]{Methods} section, the Green function is defined for a period of 2000 years (y). Therefore, in what follows we are able to perform predictions beyond the time frame simulated in the $1pctCO2$ scenario (which is only 1000 y), thus showing very useful predictive power. 

First, we analyse the response of the globally averaged near-surface temperature ($T_{2m}$) on short and long time scales. We then focus on two key aspects of the large-scale ocean circulation, namely the Atlantic Meridional Overturning Circulation (AMOC) and the  Antarctic Circumpolar Current (ACC), and show that we can achieve excellent skill in predicting the the slow modes of the climate response. We also look into the global ocean heat uptake (OHU) \cite{Exarchou2015}. A non-vanishing OHU indicates the presence of a global net energy imbalance. In current conditions, the ocean is well-known \cite{VonSchuckmann2016} to absorb a large fraction of the Earth's energy imbalance due to global warming  and to store it through its large thermal inertia, up to time scales defined by the deep ocean circulation. Finally, we prove the validity of our approach for predicting the change in the surface temperature in the North Atlantic, where the ocean deep water formation takes place. This region features a complex interplay between radiative exchanges and meridional energy transport, thus being particularly sensitive to the strength of the forcing and the changes in the large-scale circulation.

\section{Results}\label{sec:3}


\subsection{Global Mean Surface Temperature}\label{sec:3.1}

\begin{figure}[tbhp]
\centering
\includegraphics[width=\linewidth]{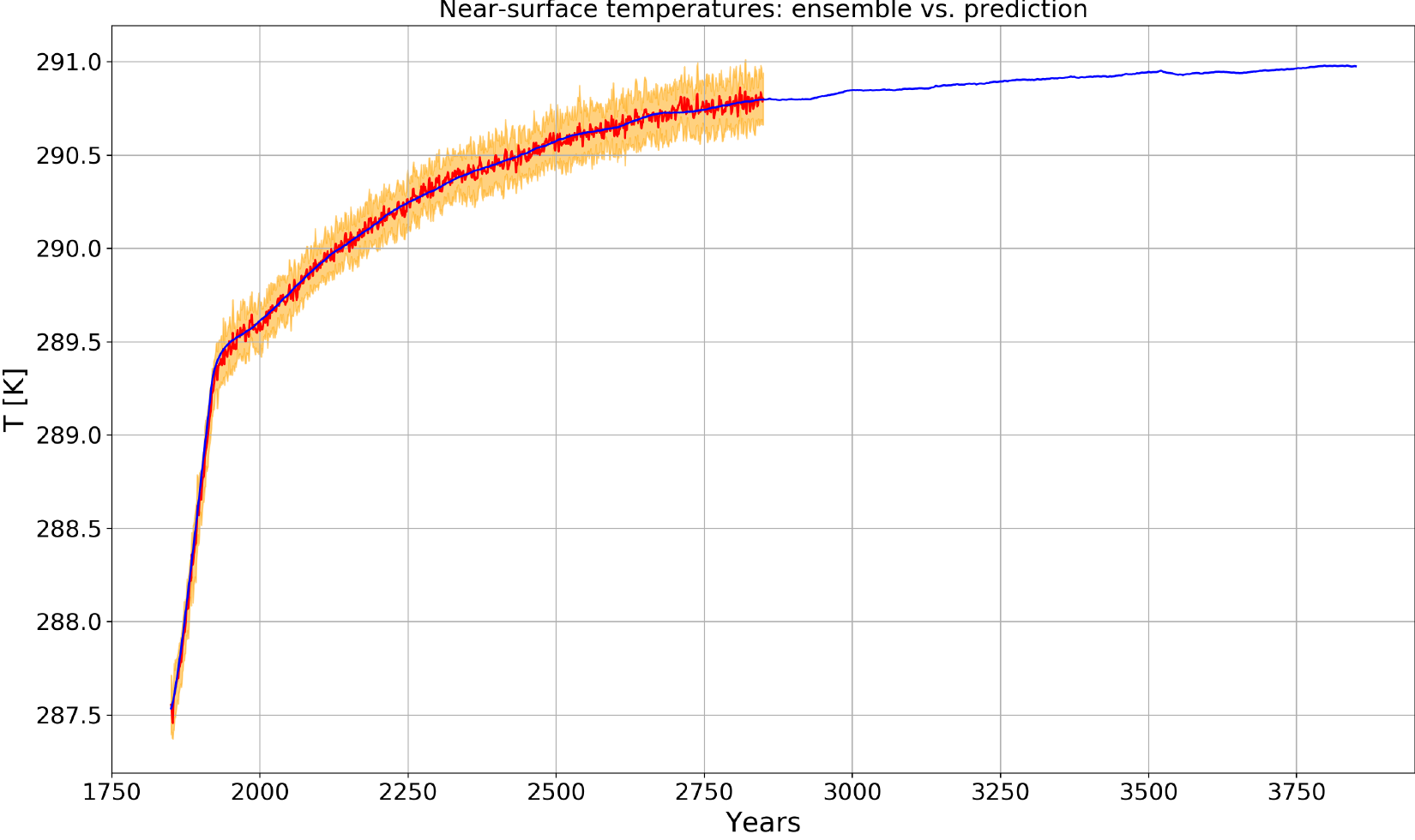}
\caption{Comparison between the simulated $T_{2m}$ in the $1pctCO2$ scenario (thick red line with ensemble mean uncertainty) and the prediction  performed using the linear Green function in Fig. \ref{figs1} (thick blue).}
\label{fig1}
\end{figure}

Figure \ref{fig1} shows that the change of  $T_{2m}$  under the $1pctCO2$ scenario {\color{black}is predicted with very good accuracy through} response theory. The prediction is accurate both for the fast (first 70 y) and the subsequent slow response. The time pattern of temperature change  indicates that the contribution of the fast feedbacks saturates after few decades, and the slow modes dominate the response for the rest of the period. The warming goes on for multicentennial scales, in a way that is not captured at all by models featuring a non-dynamic ocean \cite{Ragone2016,Lucarini2016}. The importance of the slow modes of climate response, associated with the oceanic thermal inertia, can be quantified considering the ratio between the transient climate response ($TCR$) and the equilibrium climate sensitivity ($ECS$); see \hyperref[met]{Methods}. Here we have $TCR/ECS\approx0.5$ ($ECS\approx 3.5 K$), which is much smaller than what found ($\approx 0.85$) by {\color{black}{Ragone et al. }}\cite{Ragone2016}, indicating a very prominent role of the slow modes of variability. The prediction obtained via response theory shows the establishment of steady state conditions for times larger than 1000 y. 

The Green function - Fig. \ref{figs1} - provides information on the time scales of the response. As a result of  the presence of slow oceanic time scales, the Green function  significantly departs from a simple exponential relaxation behavior, which is sometimes adopted to describe the relaxation of the climate system to forcings \cite{Hasselmann1993,Held2010}. {\color{black} The idea of defining a general Green function as a sum of multiple or infinite\cite{Lucarini2018JSP} exponential functions with different timescales, generalising Hasselmann's ideas - the so- called pulse-response method - and along the lines of previous studies \cite{Joos2013,Millar2017,Zappa2020} is beyond the scope of our analysis and will be investigated further in a future work.} In our case, after a fast decrease for short time scales, the Green function tends to zero at a much slower pace in for times longer than 100 y, in agreement with what reported by Held et al. \cite{Held2010}. 

\subsection{Atlantic Meridional Overturning Circulation}\label{sec:3.2}

\begin{figure}[tbhp]
\centering
\includegraphics[width=\linewidth]{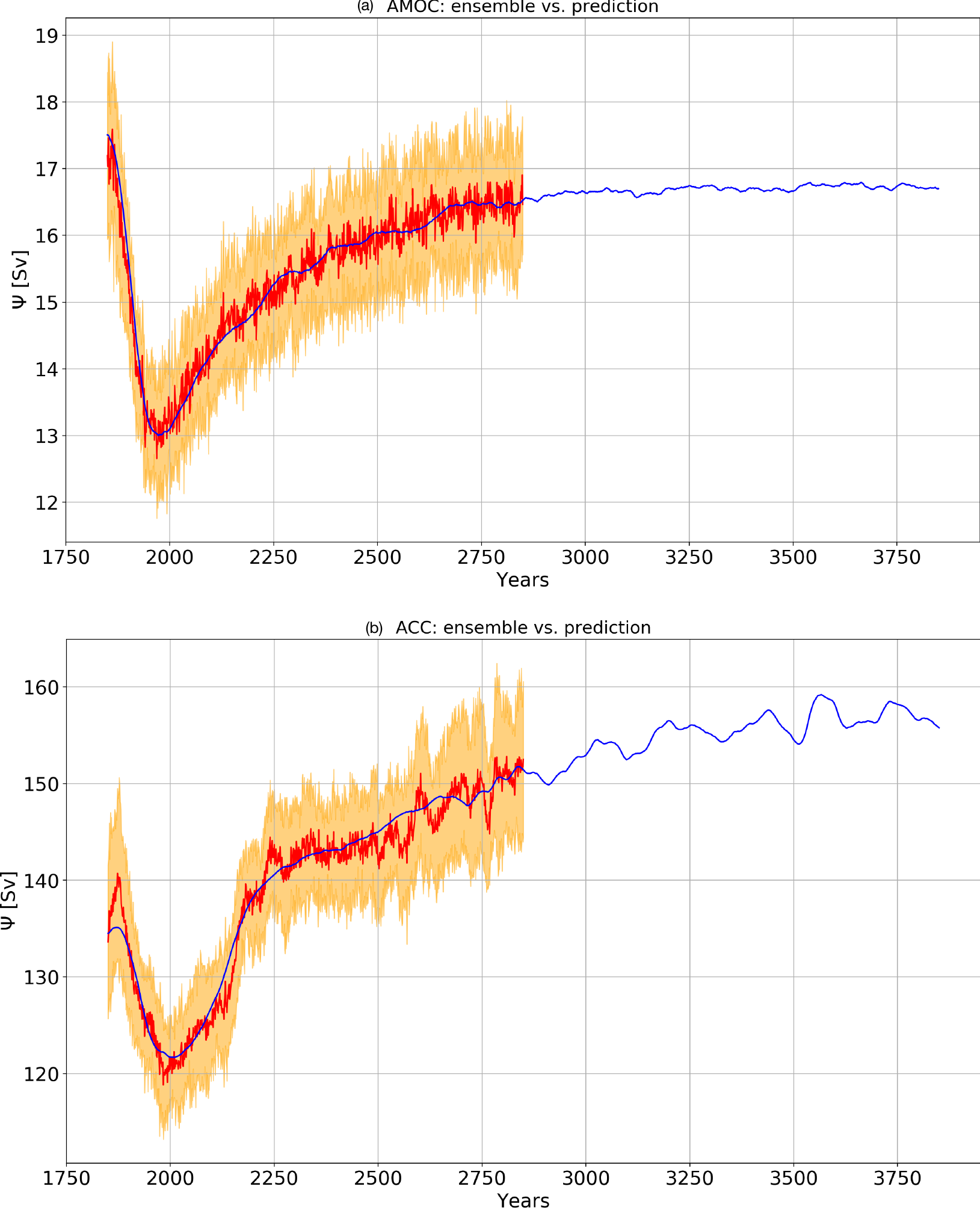}
\caption{Same as Fig. \ref{fig1}, for (a) AMOC at 26N (in $Sv$) and (b) ACC (in $Sv$). The predictions are performed using the linear Green functions shown in Fig. \ref{figs2}.}
\label{fig2}
\end{figure}

The AMOC is strongly influenced by buoyancy perturbations in the Atlantic basin \cite{Kuhlbrodt2007}. It is relevant at climatic level because it encompasses about 25\% of the total (atmospheric+oceanic) meridional heat transport \cite{Hirschi2003}. The time series of annual mean AMOC strength in the $1pctCO2$ scenario is shown in  Fig. \ref{fig2}a. The  AMOC strength undergoes a decrease by about 30\%, reaching its minimum in about 150 y. Successively, the AMOC slowly recovers. 

The prediction of the AMOC change obtained via response theory captures very well the ensemble mean of the time evolution for the $1pctCO2$ in the first 1000 y. The corresponding Green function is shown in the inset of Fig. \ref{figs2}a. On short time scales, we have a reduction of AMOC, as a result of the negative value of the Green function. On longer time scales ($>$100 y), a negative feedback acts as a a restoring mechanism, associated with a positive sign in the Green function. The presence of fast (meaning here decadal) response associated with the GHG forcing has already been found in other models \cite{Xu2019,Zickfeld2013}, and is most likely related to the timescales of the sea-ice melting, consistently with paleoclimate simulations of the last interglacial climate with prescribed freshwater influx from reconstructed sea-ice melting \cite{Loutre2014}. The slow recovery of the AMOC might be understood {\color{black} as a} heat \cite{Rahmstorf1996MW,Scott1999,LucariniStone2005} {\color{black} and freshwater \cite{Jackson2013} advection feedback}.

In the 1001-2000 y period, response theory shows that a steady state is progressively reached over multi-centennial scales. The newly established AMOC is significantly weaker than the unperturbed AMOC, although a large ensemble spread is found. This is consistent with simulations obtained from higher resolution versions of the same model \cite{Li2013}, intermediate-complexity models \cite{Zickfeld2013} and other fully coupled models inclusive of an interactive carbon cycle \cite{Zickfeld2012}. 

\subsection{The Antarctic Circumpolar Current}\label{sec:3.3}

The ACC is by far the strongest large-scale oceanic current and its role in the general circulation is two-fold. On one hand, it 
isolates Antarctica from the rest of the system, being associated with a very strong zonal circulation in the Southern Ocean. 
On the other hand, although eminently wind-driven, it marks the area of outcropping of deep water occurring at the southern flank of the subtropical gyre, as part of the global-scale overturning circulation.

The Green function is shown in the inset of Fig. \ref{figs2}b. We find that the initial strengthening of the ACC can be associated with an increase in surface zonal wind stress (not shown here). Such a surface forcing determines an enhanced Eulerian mean ACC transport, consistently with previous low resolution simulations \cite{Fyfe2006}. On decadal scales, we have a loss in the correlation between wind stress and ACC, corresponding to the Green function turning negative after about 30 y. Beyond these time scales, we have {\color{black}{time-wise coherent}} response of the AMOC and ACC, {\color{black}{underlying the}} response of the global ocean circulation. Other models \cite{Meijers2012,Heuze2015} also feature such a behavior on intermediate time scales, consistently with the idea that the two circulations are related via the thermal wind balance \cite{Marshall2017}.

Figure \ref{fig2}b shows that the prediction of the ACC strength evolution in the $1pctCO2$ scenario is rather accurate for the first 1000 y, except for an underestimation of the positive short-term response, which is smoothed out. This points to an insufficient ability of response theory in representing 
the complex coupling between surface wind stress and downward momentum transfer. {\color{black}{Furthermore,}} we observe the presence of a strong variability (on decadal time scales) of the predicted signal. This might result from either the small ensemble size or, more interestingly, could be the signature of the natural variability, {\color{black}  encoded by a Ruelle-Pollicott pole \cite{Chekroun2014,Lucarini2018JSP,Ghil2019}; see discussion in Sect. \ref{responseclimatechange}.} 
 
Note that in the 1001-2000 yrs period the ACC reaches an approximate steady state a bit later than the  AMOC, possibly as a result of having a larger inertia, consistently with the different depth scales of the two currents \cite{Koltermann2011,Marshall2017}. The AMOC maximum overturning depth scale is indeed located at about 1 km, whereas the outcropping in the Southern Ocean is related to isopycnal surfaces reaching much deeper. This has profound implications for setting the time scales of the ACC and AMOC response. The propagation of deep water formation anomalies in the Northern Hemisphere is in fact mediated by Kelvin waves in the Northern Atlantic, whereas much slower interior adjustment through Rossby waves communicates the anomaly to the Southern Ocean \cite{Johnson2002}. 

\subsection{Ocean heat uptake}\label{sec:3.4}

\begin{figure}[tbhp]
\centering
\includegraphics[width=\linewidth]{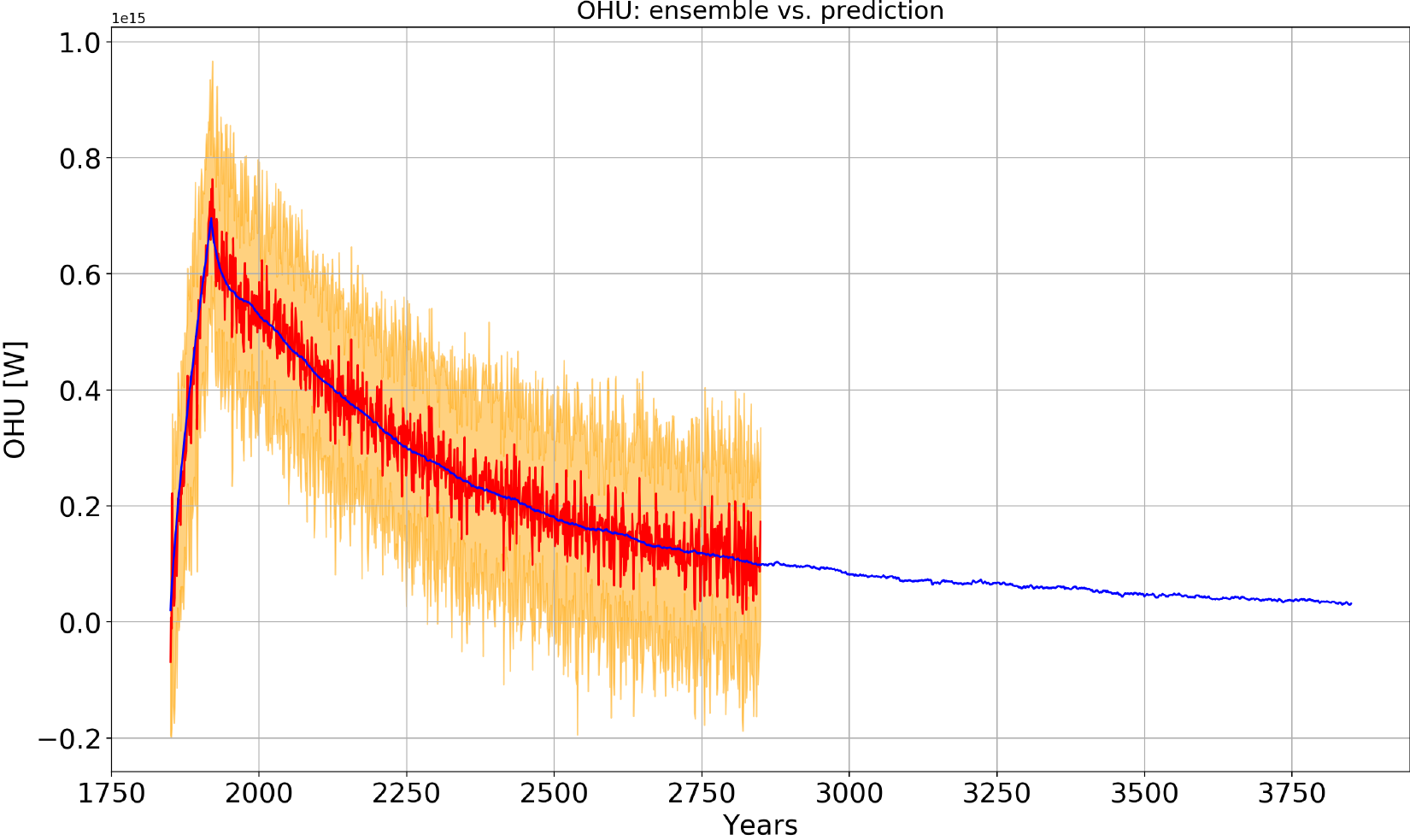}
\caption{Same as Fig. \ref{fig1} for the OHU (in $W$). The prediction is performed using the linear Green function shown in Fig. \ref{figs3}.}
\label{fig3}
\end{figure}

Looking at the 2000 y of prediction in Fig. \ref{fig3}, we notice that response theory accurately predicts the response at all time scales. The linearity of the OHU increase in the 70 y of integration comes from the convolution of the singular component of the corresponding Green function with the ramp, see the \hyperref[met]{Methods} section. After the CO$_2$ concentration stabilizes, the OHU decreases towards vanishing values. In the last 1000 y, response theory predicts a further decrease in the OHU down to a value of the order of {\color{black}$\approx 5 \times 10^{13} W$. As the  climate system as a whole relaxes towards the newly established {\color{black}energetic} steady state through the negative Planck feedback, each of its subcomponents go through a process of relaxation. What we portray in Fig. \ref{fig3} after year 1820 is the relaxation of the slowest climatic component, namely the global ocean.}. The remaining imbalance at the end of the prediction can also be interpreted as either resulting from the ultra-long time scales required for reaching rigorous steady state conditions, or as the signature of a model energy bias, associated with non-vanishing energy budget at steady state, see discussion and Fig. 1 in {\color{black}{Lucarini and Ragone }}\cite{Lucarini2011b}.

\subsection{The North Atlantic cold blob}\label{sec:3.5}

\begin{figure}[tbhp]
\centering
\includegraphics[width=\linewidth]{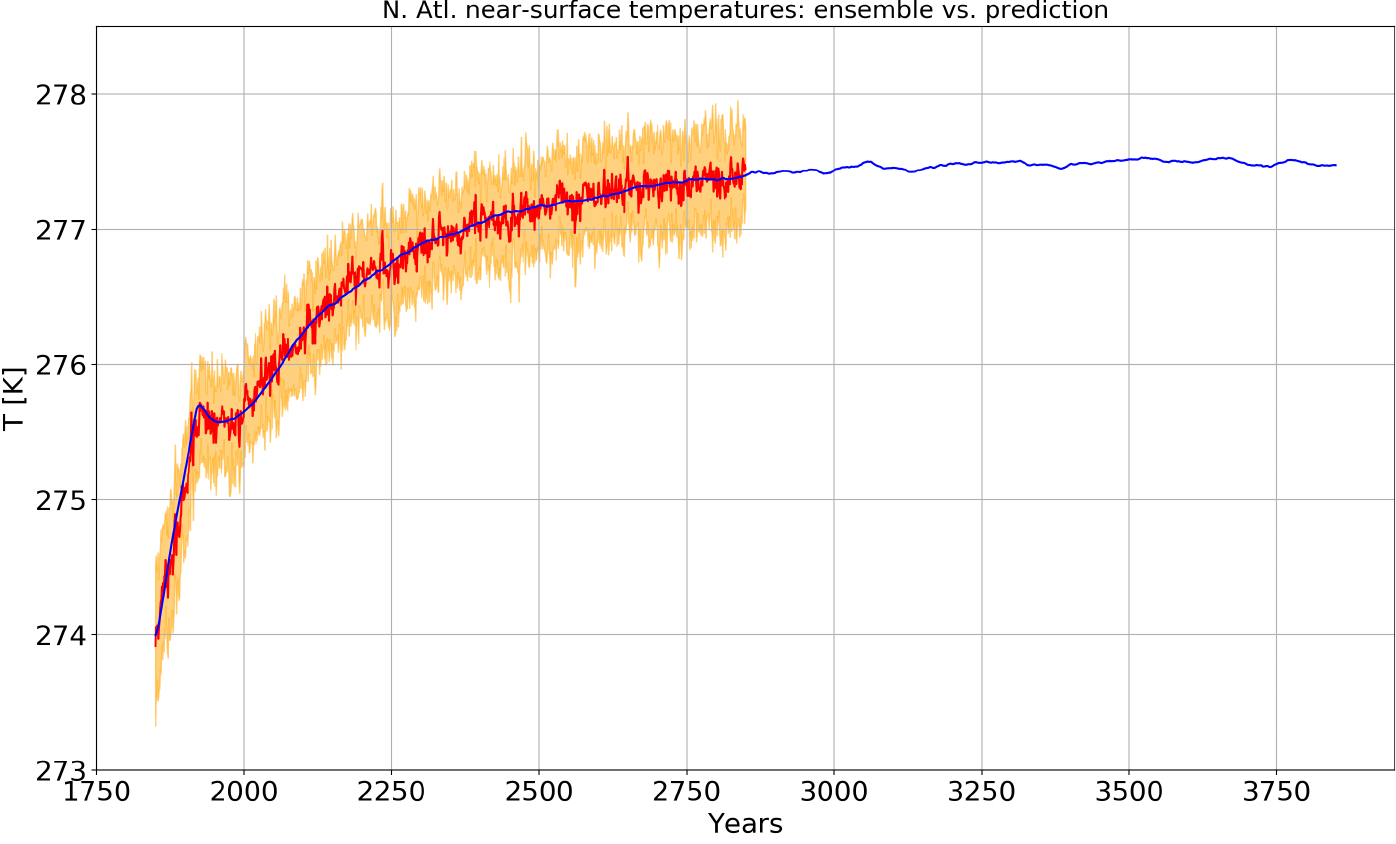}
\caption{Same as Fig. \ref{fig1} for the surface temperature in the North Atlantic (in K). The prediction is performed using the linear Green function shown in Fig. \ref{figs4}.}
\label{fig4}
\end{figure}

\begin{figure}[t]
\centering
\includegraphics[width=\linewidth]{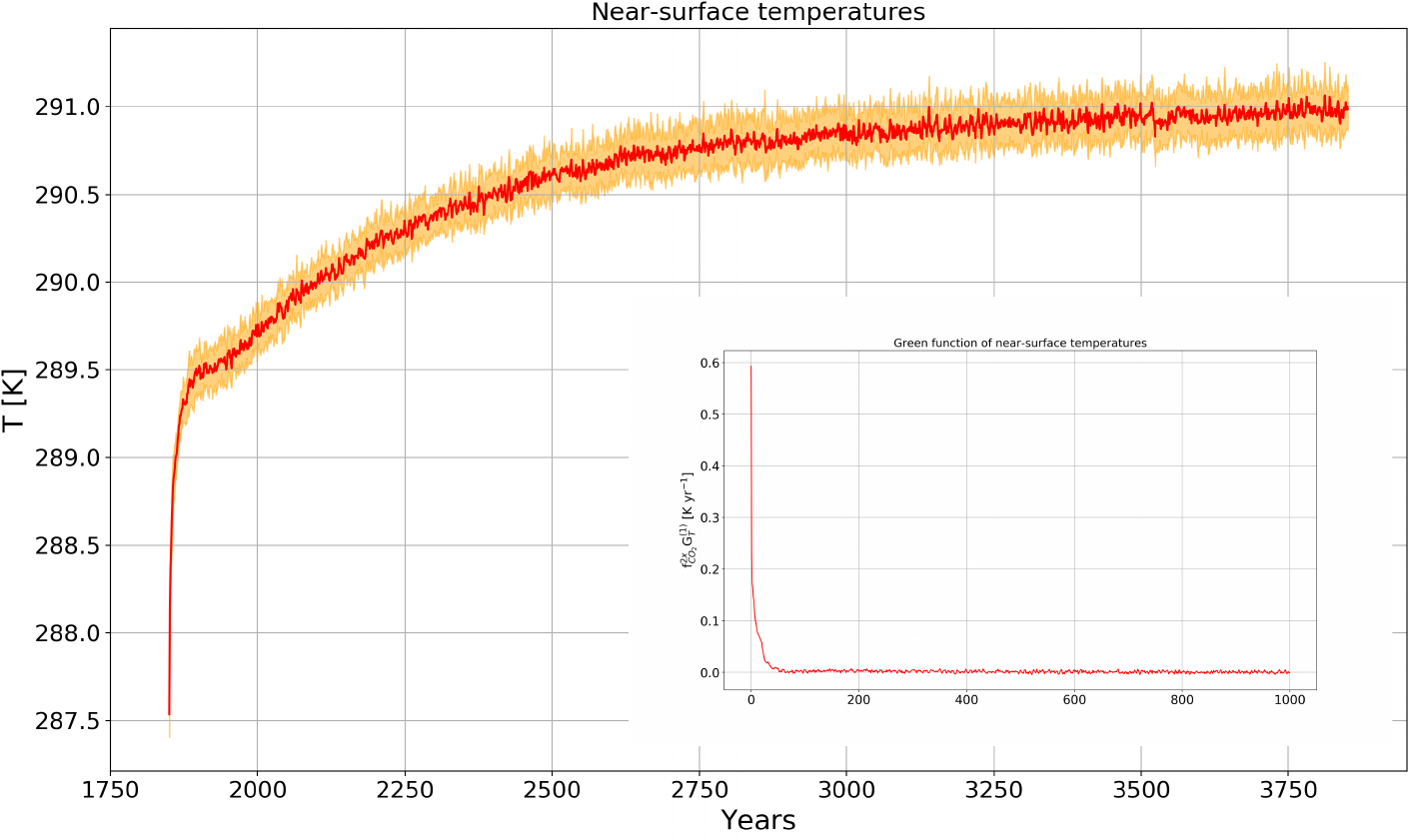}
\caption{Time series evolution of global mean $T_{2m}$ (in K) for  $2xCO2$ . The thick line the annually averaged ensemble mean, the shaded areas denote the $1\sigma$ ensemble range. The inset shows the first 1000 y of the linear Green function for global mean $T_{2m}$ (in $K yr^{-1}$), computed from the ensemble mean of the  $2xCO2$  experiment.}
\label{figs1}
\end{figure}

\begin{figure}[tbhp]
\centering
\includegraphics[width=\linewidth]{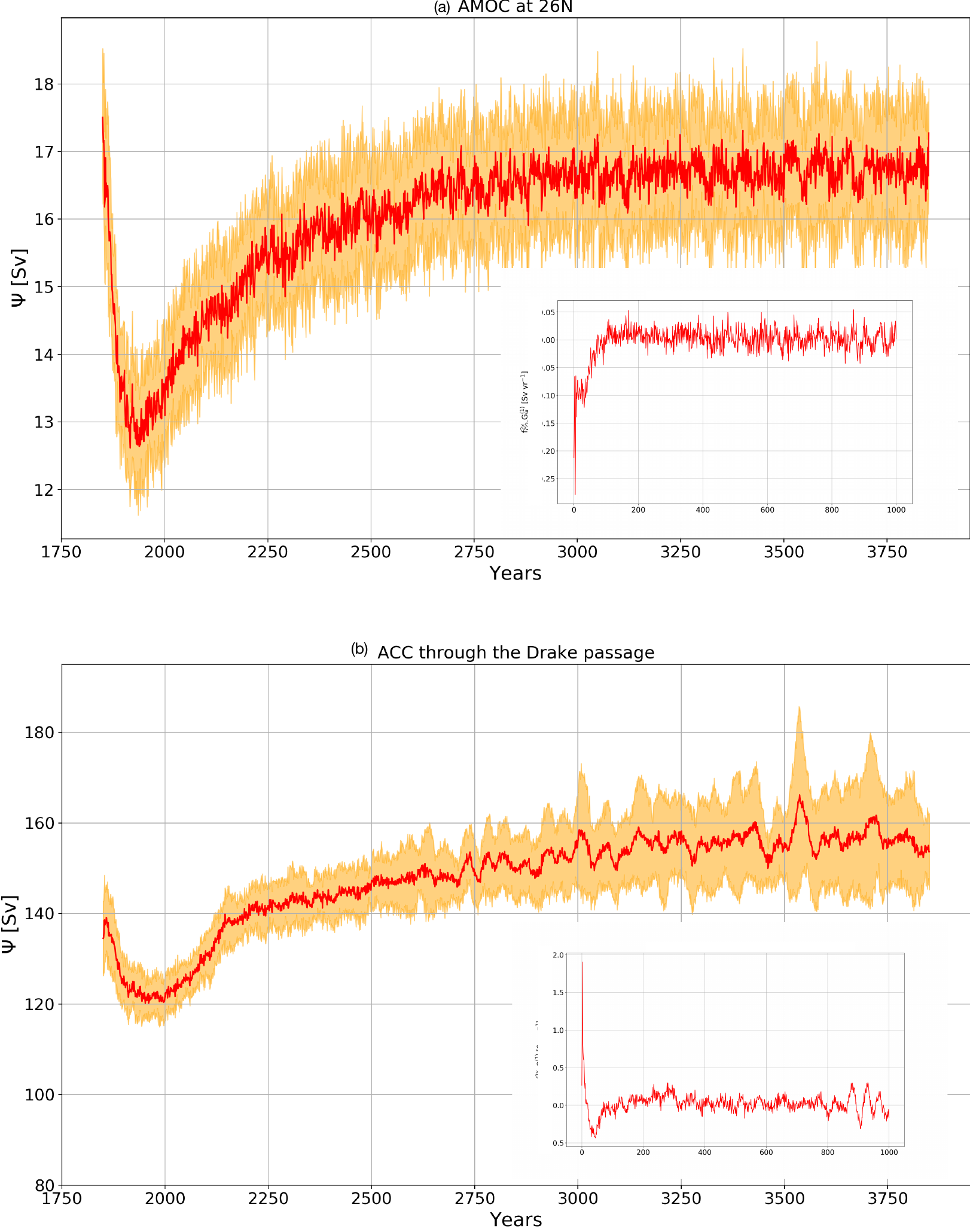}
\caption{Same as in Fig. \ref{figs1}, for (a) AMOC at 26N (in $Sv$) and (b) the ACC through the Drake passage (in $Sv$). The linear Green functions are in Sv yr$^{-1}$).}
\label{figs2}
\end{figure}

Finally, we study the surface temperature response for a domain covering the North Atlantic (
between 26$^\circ$W  and lon 53$^\circ$W and latitude between 53$^\circ$N and 69$^\circ$N), thus including the areas where the deep water formation occurs. 
The region is identified as a peculiar spot for the effects of the GHG forcing, since the sea-ice melting has been hypothesized to delay the surface warming of this area compared to surrounding regions, through a weakening of the overturning circulation \cite{Rahmstorf2015}. Indeed -  see Fig. \ref{fig4} - the surface warming over the North Atlantic region  is remarkably different from the behavior of the rest of the extratropics, which features a time dependent response (not shown here) similar in shape but somewhat amplified with respect to the global mean depicted in Fig. \ref{fig1}. Indeed, a long-lasting plateau - a hiatus in the temperature increase of more than 100 y - is observed around the end of the CO$_2$ increase ramp in the North Atlantic. The plateau is well captured by response theory, and comes in agreement with  the AMOC weakening \cite{Rahmstorf2015} predicted in Fig. \ref{fig2}a. This result is non-trivial, given that such local response results from an interplay of local factors and, as mentioned, the response of the large-scale oceanic circulation. This hints at the potential of using response theory to identify  global quantities that can be used as predictors for the response of local observables \cite{Lucarini2018JSP}.

\section{Discussion and Conclusions}
\label{sec:Concl}

We have shown here that response theory is a valuable tool for predicting crucial aspects of the response in the climate system to a prescribed forcing. All relies on obtaining the observable-dependent Green function from model simulations. 
The Green function allows one to deal with a continuum of time-dependent forcings, beyond the standard use of reference scenarios. Our findings provide guidance to the climate modellers' community on how to set up climate change experimental protocols, minimising the need for computational resources. {\color{black}This is especially relevant when one wants to study the overall impact and individual effects of multiple climatic forcings or investigate geoengineering options.} 
 
We have made here use of a fully coupled climate model, highlighting the slow components of the response associated with the oceanic modes of variability. The presence of a vast range of active time scales in the system makes the prediction of the response theoretically challenging and practically extremely relevant. {\color{black}Compared with previous contributions to the literature that have attempted with a good degree of success the prediction of changes in slow climatic fields using some form of linear regression, our approach is mathematically more robust as it derives directly from basic results in non-equilibrium statistical mechanics.} 

\textcolor{black}{We stress that response formulas based on Ruelle's theory are rigorously valid, but only when considering ensemble averages. On one side, this clarifies the limits of validity of previous attempts to apply similar formulas to individual model runs. On the other side, it indicates the importance to consider large ensemble strategies when planning major computational efforts for climate change projections.} 

The predicted changes in the AMOC and ACC feature clearly distinguishable fast and slow regimes of response. The former is essentially different in the two branches of the global ocean circulation, being the ACC subject to the effect of surface wind stress anomalies (substantially underestimated, compared to actual simulations). The latter is found to be well correlated between AMOC and ACC, as a signature of the forced response of the global ocean circulation circulation,  which they are both part of. {\color{black}{Coherently}} with previous findings \cite{Koltermann2011,Marshall2017}, the ACC reaches a steady state much later than the AMOC. The plateau in near-surface warming over the North Atlantic is also related to the initial slowdown of the AMOC, and contrasts with the regular increase of the surface temperature we find globally (and predict accurately).

We remark that the time-dependent transient and long-term {\color{black}evolutions} resulting from the CO$_2$ increase are qualitatively different for the various climatic observables investigated here. Nonetheless, in all considered cases response theory successfully predicts the time-dependent change.

{\textcolor{black}{Ruelle's response theory provides a relatively simple yet robust and powerful set of diagnostic and prognostic tools to study the response of climatic observables to external forcings. {\color{black}The availability of a large number of ensemble members allows for constructing more accurate Green functions and for studying effectively the response of a broader class of climatic observables. }
The approach proposed here could be extremely useful to inform the planning of major computational efforts for climate projections, putting such endeavours on firm theoretical grounds and optimizing the use of computational resources.}}

We have dealt here with a forcing due only to changes in the CO$_2$ concentration. This means that the pattern of the forcing is determined by heating rate associated with a spatially homogeneous CO$_2$ mixing rate. Within the linear  regime, different sources of forcings can be treated independently.
The next step is to investigate other nontrivial forcings (e.g. aerosol forcings, land-use change, land glaciers location and extension). {\color{black}As an example}, response theory has been proposed as a tool for framing geoengineering strategies and understanding its limitations \cite{bodai2018geo}. 

\textcolor{black}{Response theory provides a powerful formalism to tackle different problems related to the concepts of feedback and sensitivity. A promising application is the definition of functional relations between the response of different observables of a system to forcings, in the spirit of recent applications (see, e.g. {\color{black}Zappa et al.} \cite{Zappa2020}). This would allow to treat comprehensively the concept of feedback across different time scales and define causal links between  variables \cite{Lucarini2018JSP}. On a different line, one of the most promising future applications will be given by the synergy between the formalism of response theory and the recently proposed theory of emergent constraints \cite{Cox2018,Hall2019}. The combination of the two approaches could lead to much needed insights on the climate response to forcings.} 

{\color{black} {\color{black}Additionally,} along the lines of the pulse-response approach, it is in principle possible to try to extract the characteristic time scales of the response of the system by fitting the Green functions of the considered observables as a weighted sum of (in principle, infinitely many) exponential functions. As explained in Refs. \cite{Lucarini2018JSP,Tantet2018}, the time scales of the exponential functions are \textit{the same} for all observables. Instead, the weight of each exponential contribution does depend on the observable of interest, with the response of rapidly equilibrating observables being dominated by the fast time scales, {\color{black}{while the opposite holds}} for observables associated with the slow components of the system. The optimal fit can be obtained through a global optimisation procedure, where the response of various different observables is simultaneously fitted to a sum of exponentials. 
We will delve into this interesting problem of inverse modelling in a separate publication.} 



Clearly, in some applications one may want to test accurately to what extent nonlinear effects are relevant, as the theory is also applicable to higher orders \cite{Lucarini2009}. 
{\color{black}Some insights into the non linear component of the response could also be obtained by appropriately combining forcings differing in sign and magnitude \cite{Lucarini2009,Gritsun2017,bodai2018geo}.} 
Nonetheless, being based on a perturbative approach, response theory (linear and nonlinear) has, by definition, only a limited range of applicability (e.g. one cannot use it to treat arbitrarily strong forcings). {\color{black}{Still}}, the non-applicability of response theory has itself fundamental implications for the knowledge of the dynamics of the system one is studying. 
{\color{black}At a tipping point \cite{lenton2008,feudel2018,Ashwin2019,Lucarini2019} (or critical transition) the negative feedbacks of a system are overcome by the positive ones and any linear Green function diverges as a result of the increase in the time correlations of the system due to a critical Ruelle-Pollicott pole \cite{Chekroun2014,Ghil2019}, signalling the crisis of the chaotic attractor \cite{Tantet2018}.} 
Instead, near a critical transition, response operators  {\color{black} do not converge} unless one considers very weak forcings \cite{Chekroun2014,Lucarini2016b}. The experimental design  provided here is thus also a clear and mathematically sound strategy for the study of conditions leading to tipping points and their role for the climate response \cite{Ghil2019,Ashwin2019} in state-of-the-art climate models.

\appendix
\section{Methods}\label{met}
\subsection{Simulations}
The analysis is based on two ensembles of simulations with Max Planck Institute  Earth System Model (MPI-ESM) v.1.2 \cite{Giorgetta2013}, using its coarse resolution (CR) version. It {\color{black} features}, for the atmospheric module ECHAM6 \cite{Stevens2013}, of T31 spectral resolution (amounting to 96 gridpoints in  longitude  and 48 in  latitude) and 31 vertical levels, for the oceanic module MPI-OM \cite{Jungclaus2013} of a curvilinear orthogonal bipolar grid (GR30) (122 longitudinal and 101 latitudinal gridpoints) with 40 vertical levels. The two ensembles, each including 20 runs, are based on two different scenarios. The first one features an instantaneous doubling in CO$_2$  concentrations (from a reference value of 280 $ppm$, characteristic of pre-industrial conditions) at the beginning of the simulations ($2xCO2$), the other one an increase in the CO$_2$ concentration at the constant rate of by 1\% per year, until the  $2xCO2$  level is reached after about 70 years; afterwards, the CO$_2$ concentration is kept constant ($1pctCO2$). The procedure for the construction of the ensemble is analogous to the protocol for CMIP5 \cite{Taylor2012} and Grand Ensemble \cite{Maher2018} experiments. A control run is performed for 2000 y with pre-industrial conditions. Each of the ensemble members is initialized from a state of the control run. The initial conditions are sampled from the control run every 100 y, in order to ensure sufficient decorrelation among the respective oceanic states (at least in the mixed layer \cite{Ganopolski2002, Wunsch2008}). The  $2xCO2$  simulations are run for 2000 y, while the $1pctCO2$ simulations are run for 1000 y with the same 20 initial conditions. As an additional check, one of the  $2xCO2$  members is prolonged for 2000 additional y, in order to investigate whether the model converges to the steady state or there is an intrinsic model drift \cite{Mauritsen2012}.

\subsection{Retrieval of AMOC and ACC}
 Typically, the large-scale circulation in the ocean is measured in terms of the mass transport across a suitably chosen section of a basin. The strength of the AMOC is computed as the vertically integrated mass weighted meridional mass streamfunction across the latitude 26.5$^\circ$ N \cite{Bryden2009}. This is a standard diagnostics of the MPI-ESM model. The ensemble average of the AMOC {\color{black} volume transport} amounts to 17.3 Sv, which is consistent with recent available measurements from the RAPID monitoring array \cite{McCarthy2015}. The ACC is roughly zonally symmetric, and its location is closely related to the isopycnal slopes in the Southern Ocean. Traditionally\cite{Meijers2012}, it has been measured in terms of the strength of the mass transport across the Drake passage. Similarly, we take the vertically integrated barotropic streamfunction difference  between the 2$^\circ \times$ 2$^\circ$ boxes centered around the 68$^\circ$ W, 54$^\circ$ S and 60$^\circ$ W, 65$^\circ$ S locations. The ensemble average of the ACC is 138 Sv, which is consistent with the multi-model mean estimate found in {\color{black}Mejers et al.} \cite{Meijers2012}, amounting to 155$\pm$51 Sv. It is also not far from the value commonly used as benchmark for the assessment of climate models (173 Sv \cite{Donohue2016}). 

\subsection{Linear response theory}
Response theories allow one to predict how the statistical properties of a system changes as a result of acting modulations in its external or internal parameters. The validity of the corresponding response formulas is heavily dependent on the hypothesis that the unperturbed system is structurally stable, i.e., roughly speaking, far from bifurcations, or, in terms of geophysical systems, from tipping points (see related discussions in a climate context \cite{Lucarini2011,Chekroun2014,Ragone2016,Lucarini2016}). Rigorous derivations of response theories have been provided for the case of deterministic \cite{Ruelle1998a,Ruelle1998b,Ruelle2009} and stochastic \cite{Hairer2010} dynamics. 
We only remark here that statistical mechanical arguments encoded by the chaotic hypothesis \cite{Gallavotti1996} (a non-equilibrium analogue of the ergodic hypothesis) indicate the feasibility of the methodology proposed here.

In this paper we follow to a large extent the approach presented in  \cite{Lucarini2016} and \cite{Ragone2016} (see also \cite{Lucarini2009,Lucarini2011}) for the study of a large ensemble of intermediate-complexity atmospheric model runs and follow the deterministic route for response theory \cite{Ruelle1998a,Ruelle1998b,Ruelle2009}. Let us consider a dynamical system described by the state vector $\mathbf{\mathit{x}}$, whose dynamics is described by the set of differential equations $\dot{\mathbf{\mathit{x}}}=\mathbf{\mathit{F(x)}}$. We add a perturbation vector field of the form $\mathbf{\mathit{\Psi}}(\mathbf{\mathit{x}},t)=\mathbf{\mathit{X}}(\mathbf{\mathit{x}})f(t)$, where $X$ is the structure of the forcing in the phase space and $f$ its time modulation. The expectation value of any observable $\Phi=\Phi(x)$ can be written as:
\begin{equation}
    \langle \Phi_f (t) \rangle = \langle \Phi \rangle_0+\sum_{n=1}^{\infty} \langle \Phi \rangle_f^{(n)}(t)
    \label{pertexp}
\end{equation}
where $\Phi_0$ is the expectation value in the unperturbed state, and the  term $\Phi_f^{(n)}(t)$ gives the $n^{th}$ order perturbative contribution. We consider here only the first order contribution $\langle \Phi \rangle_f^{(1)}(t)$. The linear correction is given by the convolution of the linear Green function with the time modulation of the perturbation:
\begin{equation}
    \langle \Phi\rangle_f^{(1)}(t)=\int d\sigma_1 G_{\Phi}^{(1)}(\sigma_1)f(t-\sigma_1)
    \label{linrespmet}
\end{equation}
where $G_{\Phi}^{(1)}$ is the linear Green function of the generic observable $\Phi$.  For ease of notation we have not indicated in equation \ref{pertexp} the dependence of the response on $X$, as in the applications considered in this paper $X$ is fixed and only the time modulation $f$ is varied. Note that  for a time modulation $f$ such that $\lim_{t\to 0} f(t)=f_0$, $|f_0|$ finite, and $f(t)=0$ if $t<0$, as in the case of $f(t)=cH(t)$, {\color{black}where $c$ is a nonvanishing constant and $H$ is the Heaviside distribution} ($H(t)= 0$ for $t\leq0$ and $H(t)=1$ for $t>0$), one typically has that $\langle \Phi \rangle_f^{1)}(0)=0$, as observed in this paper for all observables except the OHU. In this latter case, one has $\lim_{t\to0} \langle \Phi\rangle _f (t) \neq 0$ because the Green function has a singularity (in the form of a Dirac's $\delta$ contribution) for $t=0$ \cite{Lucarini2018JSP}.

{\color{black}We remark that {\color{black}in previous works} \cite{Pillar2016,Kostov2017,Johnson2018,Cornish2020,Zappa2020}, the linear prediction of the desired climate observable is {\color{black}instead} obtained by convolving time pattern of another climatic observable - the \textit{driver} - rather than the actual external forcing - with an effective transfer function - rather than the true Green function. The conditions under which climate observables can be used as both predictands and predictors have been discussed by Lucarini \cite{Lucarini2018JSP}.}

\textcolor{black}{By taking the Fourier transform of the Green function $G_{\Phi}^{(1)}$, one obtains the linear susceptibility of the observable $\chi^{(1)}_{\Phi}(\omega)$, where $\omega$ is the frequency. The susceptibility gives the frequency response to a forcing $f(t)$ as $\tilde{\Phi}_f^{(1)}(\omega)=\chi^{(1)}_{\Phi}(\omega)\tilde{f}(\omega)$, where with $\tilde{\cdot}$ we indicate the Fourier transform. The susceptibility gives a spectroscopic description of the properties of the response of the observable, and its analysis can give interesting information on the most relevant time scales and related processes that determine the response of the observable.}

\subsection{Procedure for the retrieval of the Green functions}

\begin{figure}[tbhp]
\centering
\includegraphics[width=\linewidth]{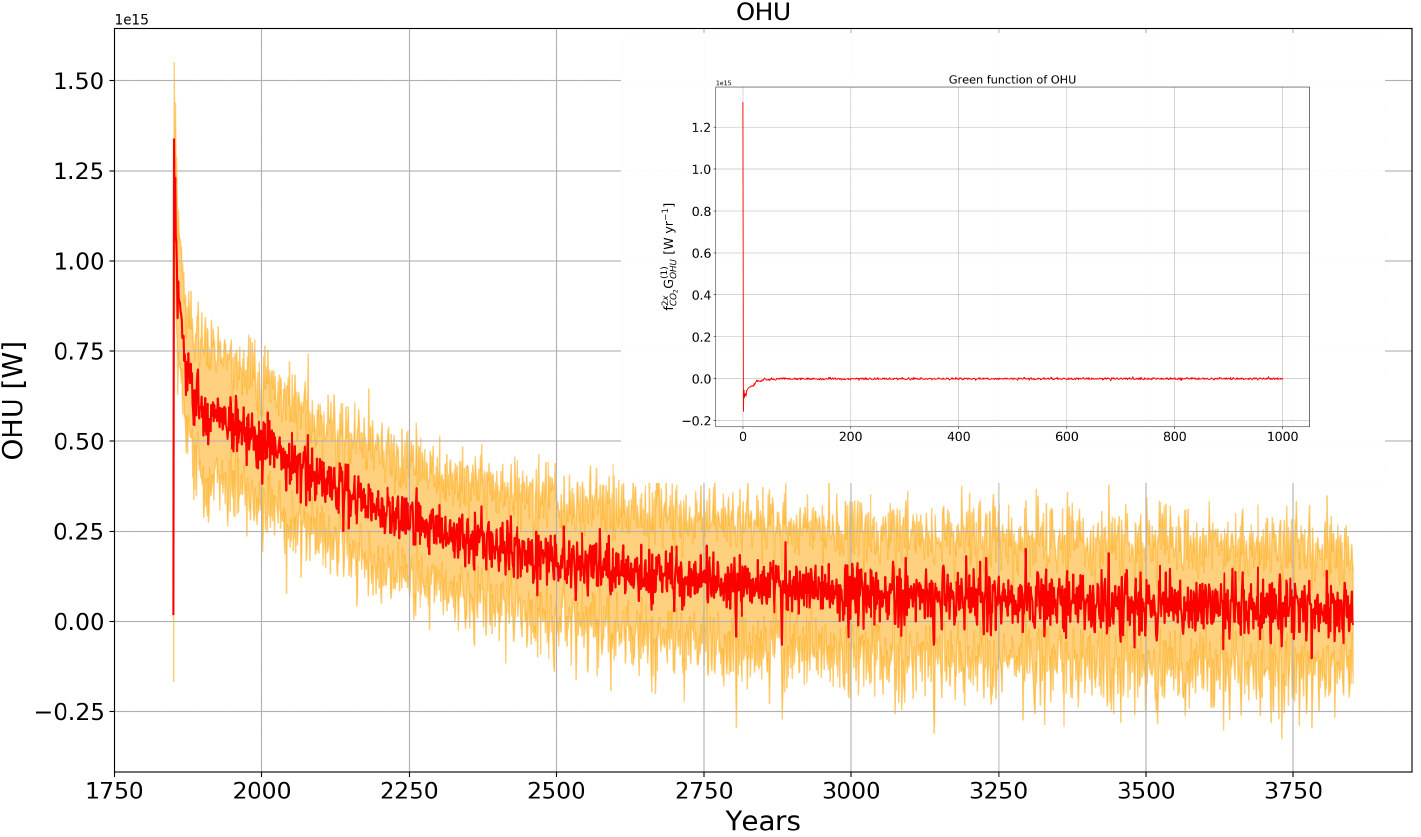}
\caption{Same as in Fig. \ref{figs1}, for OHU (in W). The linear Green function is in W yr$^{-1}$).}
\label{figs3}
\end{figure}

\begin{figure}[tbhp]
\centering
\includegraphics[width=\linewidth]{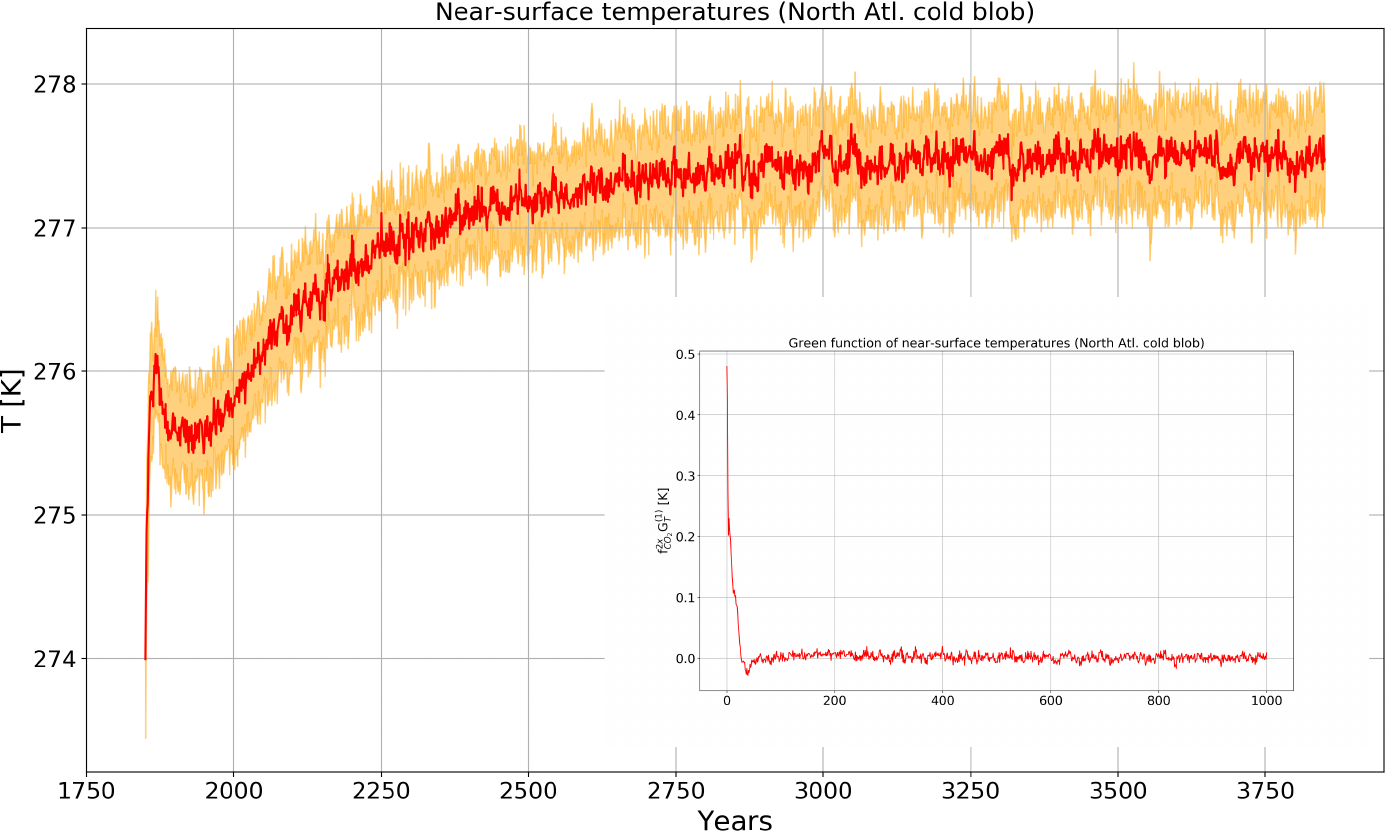}
\caption{Same as in Fig. \ref{figs1}, for the near-surface temperatures averaged in the North Atlantic region (in K). The linear Green function is in K yr$^{-1}$).}
\label{figs4}
\end{figure}

The strategy for testing the prediction of the mentioned key variables with the coupled model ensembles is as follows. First, we compute the Green function from the  $2xCO2$  experiment. {\color{black}The time variable is defined }in such a way that the instantaneous doubling occurs at $t=0$.  Hence, the time modulation of the forcing is given in this case by $f(t)=f_{2xCO_2}H(t)$, where $H(t)$ is the Heaviside function, and $f_{2xCO_2}$ is a constant depending on the amplitude of the forcing. {\color{black}Since the radiative forcing is approximately proportional to the logarithm of the CO2 concentration, such a constant is given by $f_{2xCO_2}=\log(2)$ (it would be $f_{2xCO_2}=log(p)$ if the final CO$_2$ concentration were p times as large as the initial one).} Equation \ref{linrespmet} can be thus rewritten as:
\begin{equation}
    \frac{d}{dt}\Phi_{f_{2CO_2}}^{(1)}(t)= f_{2xCO_2} G_{\Phi}^{(1)}(t)
    \label{2xresp}
\end{equation}
The outputs of the  $2xCO2$  experiments and the corresponding Green functions for the observables described above are presented in Figs. \ref{figs1}-\ref{figs4}. 

In particular, the time evolution of OHU for this scenario is shown in Fig. \ref{figs3}. The positive forcing due to the instantaneous CO$_2$ doubling leads to an instantaneous jump in the OHU, leading to an annual average value of more than 1 PW in the first year.  Equation \ref{2xresp} then suggests that the Green function $G_{OHU}^{(1)}$ has a singular behaviour at $t=0$ (cfr. \cite{Lucarini2018JSP}), while a regular behaviour is found for $t>0$, corresponding to the negative radiative Planck feedback.

For all observables, we then use the Green functions above to perform predictions for the $1pctCO2$ scenario using Eq. \ref{linrespmet}. {\color{black}Thanks to the proportionality of the radiative forcing to the logarithm of the CO$_2$ concentration, }\textcolor{black}{the time modulation of such forcing can be expressed as $f=f_{1pctCO2}r(t)$, where $r(t)$ is a ramp function (cfr. {\color{black}Lucarini et al.} \cite{Lucarini2016} and {\color{black}Ragone et al.} \cite{Ragone2016}):
\begin{equation}
    r(t)= \left\{
                \begin{array}{l}
                  \displaystyle 0 \hspace{2cm} t<0   \\\frac{t}{\tau} \hspace{2cm} 0 \leq t \leq \tau \\
                  1
                \hspace{2.3cm} t > \tau
                \end{array}
              \right.
\label{1pctforc}
\end{equation}}
where the time scale $\tau \approx 70$ y denotes the time needed to reach the doubling in the CO$_2$ concentration {\color{black} and where $f_{1pctCO2}=f_{2xCO_2}$ because at the end of the ramp the CO$_2$ concentration is doubled.} 

{\color{black}From the Green function, one could in principle compute the susceptibility and perform a spectral analysis of the properties of the response. However, the correct identification of spectral peaks in the susceptibility requires a much richer statistics than what we have available here. The reason is that, while the Green function is an integral kernel whose specific values at each $t$ are not of crucial importance per se, as it is their integrated contribution that determines the response, in the case of the susceptibility it is extremely important to make sure that the signal to noise ratio is very large \textit{at each individual value} of $\omega$. This translates into the fact that, despite the Green function and the susceptibility being strictly connected, to obtain a satisfactory estimate for the latter require a statistics orders of magnitude larger than for the former, and possibly different and dedicated numerical estimation approaches \cite{Lucarini2009,Lucarini2011,Gritsun2017}. An analysis of the susceptibility in experiments similar to what done in this work was attempted in Ragone et al. \cite{Ragone2016}, but using ten times more ensemble members. We therefore do not present an analysis of the susceptibility. While the analysis of the detailed frequency response of a climate model remains a very interesting and promising topic, it has to likely wait until experiments with at least several hundreds ensemble members will be available.}

\subsection{Equilibrium Climate Response and Transient Climate Sensitivity}\label{ECSTCR}

{\color{black}{Response theory allows to place on solid formal ground operational definitions of the sensitivity of the climate system  \cite{Ragone2016,Lucarini2016,Ghil2019}. One of the most important indicators of the global properties of the response of the system to climate change is the equilibrium climate sensitivity (ECS),} 
{\color{black}which is} the long term ($t\rightarrow \infty$) response of the observable $T_{2m}$ to an abrupt doubling of CO2 concentration \cite{IPCC2013}. Another common measure of the response is the transient climate response (TCR), which is the change in $T_{2m}$ realised in the $1pctCO2$ scenario of at the end of ramp of the CO$_2$ increase \cite{Otto2013}}. 

{\color{black}Using the formalism discussed in this paper, the ECS can be straightforwardly linked to the susceptibility, because $ECS=f_{2xCO_2}\chi^{(1)}_{T_{2m}}(0)$ \cite{Ragone2016,Lucarini2016,Ghil2019}. Additionally, the TCR can be computed as the result at time $t=\tau$ of the convolution of the Green function of  $T_{2m}$ with the forcing given in Eq. \ref{1pctforc}.} {\color{black}{Indeed, more generally, the susceptibility can be interpreted as a generalised sensitivity function. In particular, as explained in {\color{black}Ragone et al.} \cite{Ragone2016}, one can find an explicit functional relation relation between TCR and ECS {\color{black}(sometimes referred to as realised warming fraction \cite{IPCC2013})}}}:

\begin{equation}
    \frac{TCR}{ECS}=1-\int_{-\infty}^{+\infty}\frac{1+sinc(\omega \tau/2)e^{-i\omega \tau/2}}{2\pi i \omega} \frac{\chi^{(1)}_{T_{2m}}(\omega)}{\chi^{(1)}_{T_{2m}}(0)} d\omega
\label{inertia}
\end{equation}

{\color{black}where $\textrm{sinc}(x)=\sin(x)/x$. The integrand in the second term on the right hand side of Eq. \ref{inertia} gives the contribution of each time scale to the inertia of the system. Note that this approach allows for treating seamlessly the case of transient response to steeper or gentler increases of CO$_2$. However, a detailed analysis of the relationship between TCR and ECS requires an accurate estimate of the susceptibility, that as explained above is beyond the scope of this work. 

The theory of emergent constraint has been used to study the ECS \cite{Caldwell2018} and the TCR \cite{Nijsse2020} in climate models. The relationship between ECS and TCR proposed in Eq. \ref{inertia} might be helpful elucidating and better understanding  such results.}

\newpage

\section*{Acknowledgments}
Valerio Lembo was supported by the Collaborative Research Centre TRR181 "Energy Transfers in Atmosphere and Ocean" funded by the Deutsche Forschungsgemeinschaft (DFG, German Research Foundation), project No. 274762653. Valerio Lucarini was partially supported by the SFB/Transregio TRR181 project and by the Horizon2020 projects CRESCENDO (Grant no. 641816) Blue-Action (Grant No. 727852) and TiPES (Grant No. 820970). Simulations were performed with Mistral, the Deutsches Kilmarechenzentrum (DKRZ) High Performance Computing system for Earth system research (HLRE-3) supercomputer, projects No. um0005 and bu1085.

\subsection*{Author Contributions Statement}V. Le. ran the ensemble of simulations, performed the retrieval of the Green functions that were used for the predictions and analysed the results. V. Lu. proposed and directed the work and commented on the results. F. R. supported the numerical implementation of the response operators and commented on the results. All authors equally contributed to the writing of the paper.


\newpage

\end{document}